# Pressure-enhanced Interlayer Exciton in $WS_2$/$MoSe_2$ Van der Waals Heterostructure

Xiaoli Ma,[1, *] Shaohua Fu,[2, *] Jianwei Ding,[3, *] Meng Liu,[4, *] Ang Bian,[2] Fang Hong,[1, †] Jia-Tao Sun,[4, ‡] Xiaoxian Zhang,[2, §] Xiaohui Yu,[1, ¶] and Dawei He[2]

[1]*Beijing National Laboratory for Condensed Matter Physics, Institute of Physics, Chinese Academy of Sciences, Beijing 100190, China.*
[2]*Key Laboratory of Luminescence and Optical Information, Ministry of Education, Institute of Optoelectronic Technology, Beijing Jiaotong University, Beijing 100044*
[3]*CAS Key Laboratory of Standardization and Measurement for Nanotechnology, CAS Center for Excellence in Nanoscience, National Center for Nanoscience and Technology, Beijing 100190, P.R. China*
[4]*School of Information and Electronics, MIIT Key Laboratory for Low-Dimensional Quantum Structure and Devices, Beijing Institute of Technology, Beijing 100081, China*

The atomic-level van der Waals heterostructures (vdW) have been one of the most interesting quantum material systems, due to their exotic physical properties. The interlayer coupling in these systems plays a critical role to realize novel physical observation and enrich interface functionality. However, there is still lack of investigation on the tuning of interlayer coupling in a quantitative way. A prospective strategy to tune the interlayer coupling is to change the electronic structure and interlayer distance by high pressure, which is a well-established method to tune the physical properties but has rarely been used for the study of such ultrathin quantum systems. Here, we construct a high-quality $WS_2$/$MoSe_2$ heterostructure in a diamond anvil cell (DAC) and successfully tuned the interlayer coupling through hydrostatic pressure. Typical photoluminescence (PL) spectra of the monolayer $MoSe_2$ (ML-$MoSe_2$), monolayer $WS_2$ (ML-$WS_2$) and $WS_2$/$MoSe_2$ heterostructure have been observed and it's intriguing that their PL peaks shift with respect to applied pressure in a quite different way. The intralayer exciton of ML-$MoSe_2$ and ML-$WS_2$ show blue shift under high pressure with a coefficient of 19.8 meV/GPa and 9.3 meV/GPa, respectively, while their interlayer exciton shows relative weak pressure dependence with a coefficient of 3.4 meV/GPa. Meanwhile, external pressure helps to drive stronger interlayer interaction and results in a higher ratio of interlayer exciton emission with the totally disappearance of A exciton of $WS_2$ finally, indicating the enhanced interlayer exciton behavior. The first-principles calculation reveals the stronger interlayer interaction which leads to enhanced interlayer exciton behavior in $WS_2$/$MoSe_2$ heterostructure under external pressure and reveals the robust peak of interlayer exciton. This work provides an effective strategy to study the interlayer interaction in vdW heterostructures and revealed the enhanced interlayer exciton in $WS_2$/$MoSe_2$, which could be of great importance for the material and device design in various similar quantum systems.

With the emergence of two-dimensional materials, monolayer transition metal dichalcogenides (TMDs) have attracted a wide range of scientific and engineering interest due to their unique electronic structure [1-5]. Recently, they have become an ideal platform to explore fundamental scientific problems in two dimensional systems, as well as potential applications in optoelectronics because of the strong light-matter interactions [6-9]. In particular, two-dimensional (2D) TMDs with exotic properties such as an indirect to direct band gap transition in monolayers [10-12], valley-selective optical coupling [13-15], and large binding energies of excitons [16-18] have paved the way for studying the basic optical properties of the materials. Besides, with the development of diverse transfer method, vdW heterostructures composed of different materials can be randomly stacked and attached together by vdW interaction [19-22], which greatly broadened the exploration of exciton physics thus provided more opportunities for future optoelectronic applications. Many fascinating physical phenomena have been reported in various vdW heterostructures, as exemplified by transport measurements revealing Hofstadter butterfly states, fractional Chern insulators, gate-tunable Mott insulators and unconventional superconductivity, amongst other effects [23-29]. In addition, the existence of stable interlayer exciton and moiré exciton have been reported in TMD heterostructures via tuning stacking angles [30,31]. Recently, studies of $WS_2/MoSe_2$ heterostructures showed that the nearly degenerate conduction-band edges with band offset calculated to be as small as 20 or 60 meV [32,33] can promote the formation of hybridized moiré exciton due to the resonant enhancement of the hybridization strength and moiré superlattice effects [31]. An intriguing problem is how to effectively tune the electronic structure and interlayer coupling or exciton of vdW heterostructures, which will lead to the observation of new exciton behavior. As reported previously, the coupling of the hetero-interface can be tuned slightly by annealing in high vacuum [34], or by inserting hexagonal BN dielectric layers into the vdW gap [35], or by twisting angle in 2D transition metal dichalcogenide semiconductors [36]. On the other hand, high pressure has been proven to be a powerful tool to change the electronic structure and even promote phase transition of materials in the past few years [37,38]. Intriguing physical phenomena such as pressure-induced direct-indirect band gap transition [38], isostructural phase transition ($2H_c$ to $2H_a$) [37,39] and semiconductor-to-metal transition [40,41] have been successfully achieved in TMD systems. Since the optical property of TMD heterostructures is directly correlated with their electronic structures, this permits us to tune interlayer coupling and excitons in vdW heterostructures via external pressure. High pressure tuning on the interlayer excitons of vdW heterostructures has been rarely reported, which may be due to the difficulty of fabricating high-quality samples inside the high-pressure chamber. However, it's essential to investigate the efficient tuning of the optical and optoelectronic properties of vdW heterostructures by high pressure in a quantitative way. Here, we choose to study the $WS_2/MoSe_2$ heterostructure as an example by applying hydrostatic pressure via a diamond anvil cell (DAC), and observed the enhanced exciton behavior.

High-quality WS$_2$/MoSe$_2$ heterostructure has been constructed on the culet of diamond by dry transfer method. A schematic illustration of stacking process is shown in Fig. 1(a). ML-WS$_2$ is transferred onto ML-MoSe$_2$ through layer-by-layer dry transfer technique, as detailed in the supporting section (Fig. S1 [42]). Since high-quality interface is essential for charge transfer between two individual monolayer materials, the samples were annealed at 120 °C for 2 hours under vacuum. Fig. 1(b) presents the corresponding optical image of a WS$_2$/MoSe$_2$ heterostructure on a diamond substrate. The layer number of the interested MoSe$_2$ and WS$_2$ region has been identified by the contrast of the optical image, Raman spectrum, and PL spectrum. Fig. 1(c) shows the Raman characteristic peaks of the out-of-plane A$_{1g}$ modes for ML-MoSe$_2$ (at 240 cm$^{-1}$), the out-of-plane A$_{1g}$ and in-plane E$_{2g}$ mode for ML-WS$_2$ (at 417 and 353 cm$^{-1}$), respectively, and the heterostructure region clearly shows these three distinguishable modes.

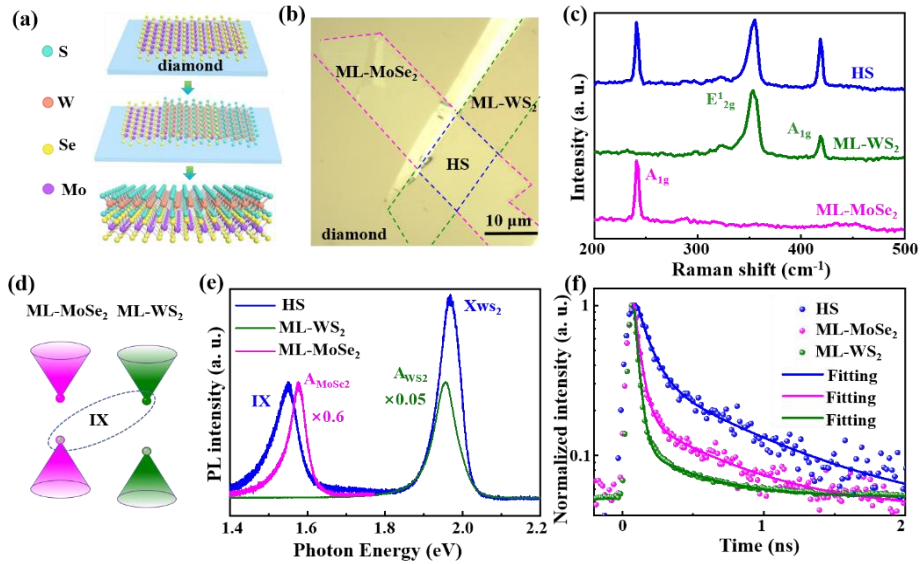

FIG. 1. Construction and optical characterization of a ML-WS$_2$/ML-MoSe$_2$ heterostructure. (a) Schematic diagram of a vertical WS$_2$/MoSe$_2$ vdW heterostructure (HS). (b) Optical micrograph of the WS$_2$/MoSe$_2$ heterostructure, fabricated by dry transfer method. ML-MoSe$_2$, ML-WS$_2$ and heterostructure regions are marked by pink, green and blue dashed lines, respectively. (c) Raman spectra of ML-MoSe$_2$, ML-WS$_2$ and WS$_2$/MoSe$_2$ heterostructure, respectively. (d) Schematic illustration of the type-II band alignment of the WS$_2$/MoSe$_2$ heterostructure and the formation of interlayer excitons. (e) PL spectra of individual monolayer and the heterostructure. The new PL peak at 1.548 eV indicates the formation of interlayer excitons (IX). (f) Time-resolved PL decay curves for ML-MoSe$_2$, ML-WS$_2$ and WS$_2$/MoSe$_2$ heterostructure.

To investigate the interlayer electronic coupling and interlayer exciton in our heterostructure, we performed micro-PL measurements and the PL spectra of ML-MoSe$_2$, ML-WS$_2$, and the WS$_2$/MoSe$_2$ heterostructure were presented in Fig. 1(e). The monolayer samples showed emission peaks at ~1.577 eV and ~1.955 eV, which corresponds to the direct inter-band K-K transition of A exciton of ML-MoSe$_2$ (A$_{MoSe_2}$) and ML-WS$_2$ (A$_{WS_2}$), respectively. For the heterostructure region, an additional emission peak at ~1.548 eV that has lower energy than A exciton of MoSe$_2$ is observed, beyond the emission peak of the A exciton of WS$_2$ (X$_{WS_2}$). According to recent report [31,43], this new peak can be assigned to interlayer exciton (IX) of WS$_2$/MoSe$_2$ heterostructure, which is formed by the recombination of electrons and holes from different layers. At the same time, the PL intensity of WS$_2$ has been quenched about 7 times, indicating that masses of the holes generated in WS$_2$ were transferred to MoSe$_2$ before radiative recombination. It is easy to understand this phenomenon because the type-Π band alignment [44,45] of WS$_2$/MoSe$_2$ heterostructure, depicted in Fig. 1(d), will facilitate the occurrence of charge transfer between the two layers. This type-Π band alignment not only provides direct channel for the interlayer coupling, but also confirms the interlayer nature of the observed peak at ~1.548 eV . The absence of A exciton peak of MoSe$_2$ from the overall PL spectrum is consistent with the previous reports [31,43], suggesting effective interlayer charge transfer across this heterostructure region.

In order to further confirm the formation of interlayer exciton, we performed time-resolved photoluminescence (TRPL) measurement at isolated WS$_2$, MoSe$_2$ and WS$_2$/MoSe$_2$ heterostructure regions, respectively. For all areas, the PL decays are fit by a two-component exponential decays, namely a fast decay τ$_1$ and a slow decay τ$_2$. From the data shown in Fig. 1(f), we obtain a fast decay lifetime of 32 ps and 44 ps for WS$_2$ and MoSe$_2$, respectively, which is similar with the transient absorption measurements of other monolayer TMD materials [46]. The fast decay component is attributed to exciton recombination, which takes up percentage of higher than 98% (Table S1 [42]), indicating that the radiative recombination channel of A exciton is dominant. Furthermore, the fast decay lifetime of 93 ps for WS$_2$/MoSe$_2$ heterostructure, obtained using a 750 nm long-pass filters, is significantly longer than the decay of individual WS$_2$ and MoSe$_2$, which is attributed to the lifetime of interlayer exciton and is comparable with previously reported values [43]. The slow decay component has a lifetime of ~418 ps, 604 ps and 774 ps for WS$_2$, MoSe$_2$ and WS$_2$/MoSe$_2$ heterostructure, respectively, which could be attributed to either different localization of the charge carriers [46] or cooling off the lattice [47].

Subsequently, we investigated the pressure effect on both the intralayer and interlayer excitonic states at monolayer and heterostructure regions, respectively (Fig. 2 and Fig. 3). The experiment setup is shown in Fig. 3(a), which combined a micro-PL system with a DAC device that is capable of implementing high pressure on the sample transferred on diamond. By adjusting the DAC device manually and using the PL peak position of a ruby as a calibration of pressure, various pressures can be effectively applied on the sample. The arrangement on the right side of Fig. 3(a) accords with the most stable stack structure in theoretical calculations (2H stacking) (Fig. S2 and Table S2 [42]).

Fig. 2(a-c) show the typical PL spectra of ML-MoSe$_2$, WS$_2$/MoSe$_2$ heterostructure and ML-WS$_2$ under various pressures. When hydrostatic pressure is applied to ML-MoSe$_2$, ML-WS$_2$ and WS$_2$/MoSe$_2$ heterostructure respectively, the emission intensities of their PL peaks all decreased. Inspired by the previous reports in MoS$_2$ systems [38,48,49], the PL results for ML-WS$_2$ and ML-MoSe$_2$ here may be correlated with the pressure-induced direct-to-indirect band transition, which would decrease the PL quantum yield thus lower the emission intensities dramatically.

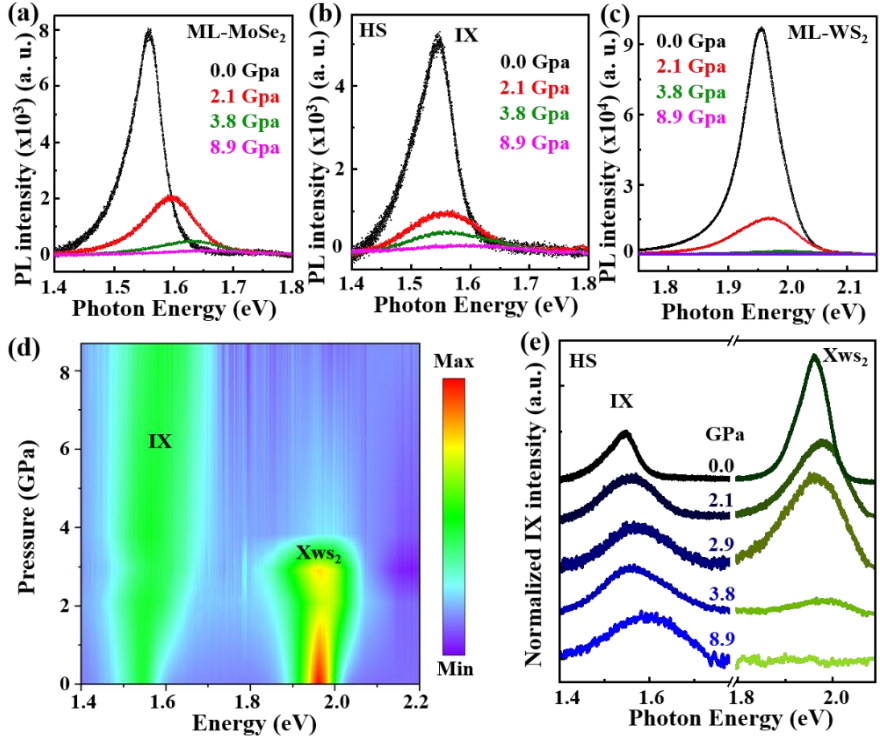

FIG. 2 (a-c) PL spectrum of ML-MoSe$_2$, WS$_2$/MoSe$_2$ heterostructure and ML-WS$_2$ under different pressures from 0 GPa to 8.9 GPa, respectively. (d-e) Normalized the intensity of interlayer exciton peaks under different pressures. As the pressure is increased, the A exciton of WS$_2$, which is the prominent emission at pressure nearly zero, is suppressed compared to the interlayer exciton (IX) emission in the heterostructure region.

In contrast to monolayer TMDs, the IX peak of WS$_2$/MoSe$_2$ heterostructure exhibited relatively more sophisticated evolution with pressure, because it not only corresponds to the electronic structure of both components but also depends on the interlayer coupling [50]. Therefore, we systematically investigated the PL spectra of WS$_2$/MoSe$_2$ heterostructure with pressure ranging from 0 to 8.9 GPa. As seen in Fig. 2(d-e), the intensity of IX and $X_{WS_2}$ (normalized by IX intensity) in heterostructure exhibits strong pressure dependence. Firstly, we observed gradual suppression of $X_{WS_2}$ peak in the PL spectrum of heterostructure compared with IX as pressure increasing. Interestingly, this $X_{WS_2}$ peak totally disappeared and only the IX peak remained in the PL spectrum of heterostructure as the pressure increasing to 8.9 GPa. This illustrates the transformation of the major PL emission from intralayer exciton to interlayer exciton in this heterostructure system. That is, the interlayer coupling of vdW heterostructure can be effectively enhanced by applying external pressure.

We further define R as the intensity ratio of IX peak and $X_{WS_2}$ exciton peak ($P_{IX}/P_{X_{WS2}}$) to trace the trend of PL peaks as pressure increasing (Table S3 [42]). Here, for the initial pressure at nearly zero, we note that the $P_{X_{WS2}}$ (exciton emission intensity) is typically 2-3 times stronger than $P_{IX}$ (interlayer exciton emission intensity), possibly due to the relatively weak interlayer coupling and indirect nature of interlayer exciton. Through statistical analysis, the ratio R is from 1:2.6 at 0 GPa to 1:0.3 at 3.8 GPa. It's notable that at 8.9 GPa only the emission of interlayer excitons was detectable in PL spectrum, indicating that interlayer exciton emission is more robust under higher pressure. Therefore, the trend of ratio R can reflect directly the tuning of interlayer electronic coupling strength under various pressures: the higher the ratio, the stronger the coupling strength. As shown on the right side of Fig. 3(a), the proportion of interlayer exciton is gradually enhanced with pressure increasing and finally totally governs the emission of heterostructure.

Besides, the peak width of interlayer exciton also shows significant change once pressure is applied (Fig. 2(d)), but remains almost unchanged as pressure increasing, which is consistent with the calculated results that a band changeover from direct to indirect transition emerging under small pressure (Fig. 4(g)). Nevertheless, specific identification of different components of interlayer exciton peak needs further experiments at lower temperature. Above all, using PL as a probe, we have achieved strong tuning of interlayer electronic coupling strength via external pressure.

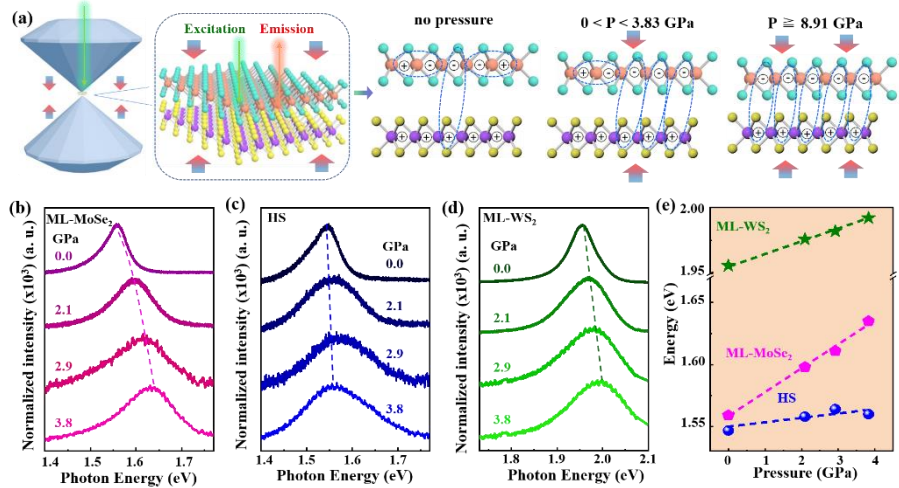

FIG. 3. Pressure engineering of electronic structure and interlayer coupling in $WS_2/MoSe_2$ heterostructure. (a) Schematic illustration of band engineering of interlayer and intralayer exciton in a vertical $WS_2/MoSe_2$ vdW heterostructure, where hydrostatic pressure is applied through a DAC device. (b-d) Pressure-dependent PL spectra of ML-$MoSe_2$, $WS_2/MoSe_2$ heterostructure and ML-$WS_2$, respectively. PL spectra of monolayer regions are individually normalized to the peak of maximum intensity. In the heterostructure region, the intensity of interlayer exciton peaks is normalized to compare the coupling strength at varying pressures. (e) Photon energies of the PL peak $A_{MoSe_2}$, $A_{WS_2}$ and IX as a function of pressure.

Not only the electronic coupling strength but also the electronic structure can be tuned by pressure, which provides us a good opportunity to study the nature of the interlayer exciton of the $WS_2/MoSe_2$ heterostructure. Since we have concluded that the electronic coupling strength was largely enhanced under high pressure. To better understand the effects of high pressure on the electronic structure, we carefully analyzed the corresponding emission peak energy of ML-$MoSe_2$, ML-$WS_2$, and the

WS$_2$/MoSe$_2$ heterostructure under pressure. As shown in Fig. 3(b) and (d), it's clear that the blue-shift of the PL peaks of both ML-MoSe$_2$ and ML-WS$_2$ were observed as the pressure increasing. This energy shift may be related to bandgap tuning by pressure [38]. The peak energy evolution of A$_{MoSe_2}$ and A$_{WS_2}$ as function of pressure is performed in Fig. 3(e) and shows a linear increasement at a rate of 19.8 meV/GPa and 9.3 meV/GPa, respectively. This behavior is consistent with previous work on ML-MoS$_2$ [38,48], which had shown that PL peak exhibited a blue shift with pressure. Moreover, previous theoretical calculation indicates that the K valley of the conduction band will shift upward under a compressive strain in agreement with our calculations [51-53]. Interestingly, comparing the pressure effect on the A exciton energies of the ML-MoSe$_2$ and ML-WS$_2$, interlayer exciton energy only shows weak pressure dependence, which was significantly different from the high-pressure response of other bilayer TMD samples. For example, the indirect inter-band transition of bilayer MoS$_2$ exhibit complex evolution) [48]. The non-sensitive behavior of the interlayer exciton energy could be due to the following two factors: 1) the conduction band edge of WS$_2$ and valence band edge of MoSe$_2$ shifts to the same direction with comparatively rate, the energy of interlayer exciton transition corresponds to them could keep almost the same value or only display weak dependence; 2) it is possible that hybridization bands [31] exist near the band edge of both components and they are not sensitive to pressure.

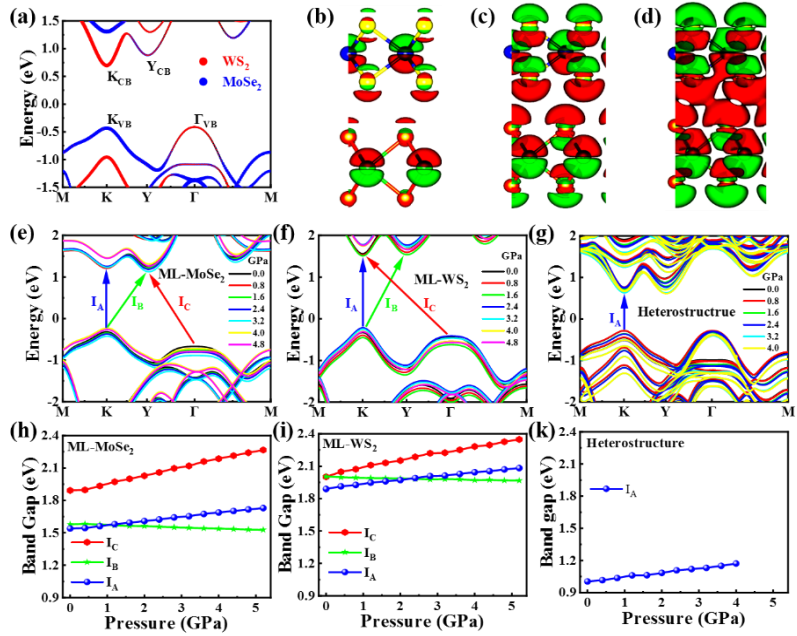

FIG. 4. Calculated band structure and differential charge density of WS$_2$/MoSe$_2$ hetero-bilayer under external pressure. (a) Calculated band structure of WS$_2$/MoSe$_2$ hetero-bilayer. The relative contribution from each layer (MoSe$_2$ and WS$_2$, denoted by the blue and red dot, respectively) is represented by the size of the symbols. (b-d) Differential charge density of WS$_2$/MoSe$_2$ hetero-bilayer when the external pressure is at 0.7 GPa, 2.0 GPa and 4.0 GPa, respectively. (Rad = positive, Green = negative, Isosurface level = 0.005). (e) and (f) Calculated band structure of monolayer MoSe$_2$ and WS$_2$ under external pressure. (g) Electronic band structures of WS$_2$/MoSe$_2$ hetero-bilayers as a function of external pressure in the range of 0-4.0 GPa. (h), (i) and (k) Calculated band gap of typical electronic band transition in the momentum space (marked in figure (e), (f) and (g)) as a function of external pressure.

To better understand the effects of high pressure on the electronic coupling and electronic structure, we conducted the density functional theory (DFT) calculation. The calculated band structure and partial charge density (Fig. 4(a) and Fig. S3 [42]) show that the valence band maximum (VBM) and conduction band minimum (CBM) of WS$_2$/MoSe$_2$ heterostructure originate from MoSe$_2$ and WS$_2$ layer, respectively. And from the differential charge density in Fig. 4(b-d), it can be seen that as the pressure gradually increases, the coupling strength gradually increases, which is in good agreement with our experimental results: at 0 GPa when the coupling strength is relatively weak, the emission of heterostructure is dominated by ML-WS$_2$; while with the increasing of coupling strength under higher pressure, the emission of interlayer exciton becomes gradually dominant; at 8.9 GPa, the emission of WS$_2$ is totally suppressed and the emission of IX totally governs the heterostructure PL spectrum. In addition, we have confirmed through DFT calculations that the band gaps of ML-MoSe$_2$ (Fig. 4(e) and (h)), ML-WS$_2$ (Fig. 4(f) and (i)) and WS$_2$/MoSe$_2$ heterostructure all change from direct band gaps to indirect band gaps with the increase of pressure. The WS$_2$/MoSe$_2$ hetero-bilayer are known to exhibit a type-II band alignment with the CBM (VBM) located at the K (K) point. In a hetero-bilayer with relative weak interlayer interaction, both intralayer and interlayer exciton emissions from individual layers can be observed, which is different from the WSe$_2$-MoSe$_2$ hetero-bilayer [50]. As interlayer interaction strength increasing, the efficient charge separation process can lead to exclusive emissions from interlayer excitons formed by electrons and holes from different layers [30]. Specifically, in free hetero-bilayer WS$_2$/MoSe$_2$ (Fig. 4(g)), the VBM is at the K point. The band gap of hetero-bilayer WS$_2$/MoSe$_2$ is very sensitive to pressure, and very small pressure can make it from direct to indirect band gap. As pressure increases, a band changeover occurs and the VBM changes from K to Γ (Fig. S4 [42]).

In addition, previous calculations show that the hybridized $d_{x^2-y^2}$ and $d_{xy}$ orbitals dominate the VBM of monolayer transition metal dichalcogenides, while the $d_{z^2}$ orbitals dominate the CBM. The similar orbital contribution from metal atoms dominates the heterostructure WS$_2$/MoSe$_2$ as well (show in Fig. S3 [42]). The orbital overlapping of the sulfur and selenium atoms is strongly dependent on the interlayer distance. When the external pressure increases, the enhanced interlayer coupling leads to strong interaction of chalcogenide atoms while leaving the negligible interaction between metal atoms. Thus, the direct interlayer hopping remains almost intact, but the indirect interlayer hopping dominates leading to the changeover between direct band gap and indirect band gap. Intriguingly, the applied pressure for the band changeover of WS$_2$/MoSe$_2$ heterostructure is much smaller than that for individual monolayer material.

In summary, we have effectively tuned the electronic structure and interlayer coupling in the high-quality WS$_2$/MoSe$_2$ heterostructure in a quantitative way, by applying hydrostatic pressure via diamond anvil cell. First, we obtained high-quality

WS$_2$/MoSe$_2$ heterostructure on diamond by dry transfer and annealing under vacuum. Both the intralayer exciton energy of ML-MoSe$_2$ and ML-WS$_2$ show clear blue shift while the interlayer exciton energy only shows weak pressure dependence, significantly different from the high pressure response of other bilayer TMD samples. For this phenomenon, we propose two possible explanations: Firstly, the conduction band edge of WS$_2$ and valence band edge of MoSe$_2$ shifts to the same direction with comparatively rate; secondly, it is possible that hybridization bands exist near the band edge of both components and they are not sensitive to pressure. Theoretical calculation reveals the enhanced interlayer interaction via the chalcogenide atoms, which is consistent with the observed enhanced ratio of IX/X$_{WS2}$ PL intensity at the heterostructure region (corresponded to enhanced interlayer exciton). Meanwhile, theoretical calculation also provides the signature of pressure induced direct-indirect band gap transition in ML-MoSe$_2$, ML-WS$_2$ and WS$_2$/MoSe$_2$ heterostructure, which could affect the PL intensity of individual monolayer and heterostructure as well. Our work has provided a good way to understand the correlation between interlayer interaction and electronic/optical properties in atomic-level vdW heterostructures, which is beneficial for electronic device design from the large quantum material family and their potential applications.

This work was supported by the National Natural Science Foundation of China (NSFC 11974088, 61875236, 61975007, 61527817, 11974045 and 11575288), the Beijing Natural Science Foundation (Grant No. Z190006), National Key Research and Development Program of China (2016YFA0202302, 2016YFA0401503, 2020YFA0308800, 2016YFA0202300 and 2018YFA0305700), the Strategic Priority Research Program and Key Research Program of Frontier Sciences of the Chinese Academy of Sciences (Grant Nos. XDB30000000, XDB33000000, XDB25000000 and Grant No. QYZDBSSW-SLH013) and the Youth Innovation Promotion Association of Chinese Academy of Sciences under Grant No 2016006.

* These authors contribute equally to this work.
Corresponding authors:
†e-mail address: hongfang@iphy.ac.cn
‡e-mail address: jtsun@bit.edu.cn
§e-mail address: zhxiaoxian@bjtu.edu.cn
¶e-mail address: yuxh@iphy.ac.cn

[1]   H. Yuan *et al.*, Nat. Phys. **9**, 563 (2013).

[2]   H. Wang *et al.*, Proc. Natl. Acad. Sci. **110**, 19701 (2013).

[3]   L. Sun *et al.*, Phys. Rev. Lett. **111**, 126801 (2013).

[4]   W. Jin *et al.*, Phys. Rev. Lett. **111**, 106801 (2013).

[5]   H. P. Komsa, J. Kotakoski, S. Kurasch, O. Lehtinen, U. Kaiser, and A. V. Krasheninnikov, Phys. Rev. Lett. **109**, 035503 (2012).

[6]   Q. H. Wang, K. Kalantar-Zadeh, A. Kis, J. N. Coleman, and M. S. Strano, Nat. Nanotechnol. **7**, 699 (2012).

[7]   S. Tongay, J. Zhou, C. Ataca, K. Lo, T. S. Matthews, J. Li, J. C. Grossman, and J. Wu, Nano Lett.

**12**, 5576 (2012).

[8] H. Yuan *et al.*, Nat. Nanotechnol. **9**, 851 (2014).

[9] D. Y. Qiu, F. H. da Jornada, and S. G. Louie, Phys. Rev. Lett. **111**, 216805 (2013).

[10] K. F. Mak, C. Lee, J. Hone, J. Shan, and T. F. Heinz, Phys. Rev. Lett. **105**, 136805 (2010).

[11] A. Splendiani, L. Sun, Y. Zhang, T. Li, J. Kim, C. Y. Chim, G. Galli, and F. Wang, Nano Lett. **10**, 1271 (2010).

[12] Y. Zhang *et al.*, Nat. Nanotechnol. **9**, 111 (2014).

[13] D. Xiao, G. B. Liu, W. Feng, X. Xu, and W. Yao, Phys. Rev. Lett. **108**, 196802 (2012).

[14] H. Zeng, J. Dai, W. Yao, D. Xiao, and X. Cui, Nat. Nanotechnol. **7**, 490 (2012).

[15] K. F. Mak, K. He, J. Shan, and T. F. Heinz, Nat. Nanotechnol. **7**, 494 (2012).

[16] K. F. Mak, K. He, C. Lee, G. H. Lee, J. Hone, T. F. Heinz, and J. Shan, Nat. Mater. **12**, 207 (2013).

[17] A. Chernikov, T. C. Berkelbach, H. M. Hill, A. Rigosi, Y. Li, O. B. Aslan, D. R. Reichman, M. S. Hybertsen, and T. F. Heinz, Phys. Rev. Lett. **113**, 076802 (2014).

[18] K. He, N. Kumar, L. Zhao, Z. Wang, K. F. Mak, H. Zhao, and J. Shan, Phys. Rev. Lett. **113**, 026803 (2014).

[19] M.-Y. Li, C.-H. Chen, Y. Shi, and L.-J. Li, Mater. Today **19**, 322 (2016).

[20] A. K. Geim and I. V. Grigorieva, Nature **499**, 419 (2013).

[21] Y. Liu, N. O. Weiss, X. Duan, H.-C. Cheng, Y. Huang, and X. Duan, Nat. Rev. Mater. **1**, 16042 (2016).

[22] K. S. Novoselov, A. Mishchenko, A. Carvalho, and A. H. C. Neto, Science **353**, aac9439 (2016).

[23] L. A. Ponomarenko *et al.*, Nature **497**, 594 (2013).

[24] B. Hunt *et al.*, Science **340**, 1427 (2013).

[25] C. R. Dean *et al.*, Nature **497**, 598 (2013).

[26] E. M. Spanton, A. A. Zibrov, H. Zhou, T. Taniguchi, K. Watanabe, M. P. Zaletel, and A. F. Young, Science **360**, 62 (2018).

[27] Y. Cao, V. Fatemi, S. Fang, K. Watanabe, T. Taniguchi, E. Kaxiras, and P. Jarillo-Herrero, Nature **556**, 43 (2018).

[28] Y. Cao *et al.*, Nature **556**, 80 (2018).

[29] G. Chen *et al.*, Nat. Phys. **15**, 237 (2019).

[30] C. Jin *et al.*, Nature **567**, 76 (2019).

[31] E. M. Alexeev *et al.*, Nature **567**, 81 (2019).

[32] C. Gong, H. Zhang, W. Wang, L. Colombo, R. M. Wallace, and K. Cho, Appl. Phys. Lett. **103**, 053513 (2013).

[33] J. Kang, S. Tongay, J. Zhou, J. Li, and J. Wu, Appl. Phys. Lett. **102**, 012111 (2013).

[34] S. Tongay *et al.*, Nano Lett. **14**, 3185 (2014).

[35] H. Fang *et al.*, Proc. Natl. Acad. Sci. **111**, 6198 (2014).

[36] K. Liu *et al.*, Nat. Commun. **5**, 4966 (2014).

[37] Z. H. Chi, X. M. Zhao, H. Zhang, A. F. Goncharov, S. S. Lobanov, T. Kagayama, M. Sakata, and X. J. Chen, Phys. Rev. Lett. **113**, 036802 (2014).

[38] L. Fu *et al.*, Sci. Adv. **3**, e1700162 (2017).

[39] Z. Zhao *et al.*, Nat. Commun. **6**, 7312 (2015).

[40] A. P. Nayak *et al.*, ACS nano **9**, 9117 (2015).

[41] A. P. Nayak *et al.*, Nat. Commun. **5**, 3731 (2014).

[42] (See Supplemental Material for (1) Methods; (2) calculated partial charge density of

WS$_2$/MoSe$_2$ hetero-bilayer; (3) PL lifetime and corresponding percentum of ML-MoSe$_2$, ML-WS$_2$ and WS$_2$/MoSe$_2$ heterostructure; (4) the top and side views of WS$_2$/MoSe$_2$ hetero-bilayer with high-symmetry stacking structures; (5) The binding energy Eb of bilayer WS$_2$/MoSe$_2$ hetero-bilayers with different stacking configurations; (6) the pressure dependent band structures of WS$_2$/MoSe$_2$ hetero-bilayer including spin-orbit coupling; and (7) the intensity ratio of IX peak and A exciton peak (P$_{IX}$/ P$_{X_{WS_2}}$) with different pressure.).

[43] F. Ceballos, M. Z. Bellus, H. Y. Chiu, and H. Zhao, Nanoscale **7**, 17523 (2015).
[44] D. A. Ruiz-Tijerina and V. I. Fal'ko, Phys. Rev. B **99**, 125424 (2019).
[45] Y. Meng *et al.*, Nat. Commun. **11**, 2640 (2020).
[46] S. Anghel, F. Passmann, C. Ruppert, A. D. Bristow, and M. Betz, 2D Mater. **5**, 045024 (2018).
[47] C. Ruppert, A. Chernikov, H. M. Hill, A. F. Rigosi, and T. F. Heinz, Nano Lett. **17**, 644 (2017).
[48] X. Dou, K. Ding, D. Jiang, and B. Sun, Acs Nano **8**, 7458 (2014).
[49] H. J. Conley, B. Wang, J. I. Ziegler, R. F. Haglund, Jr., S. T. Pantelides, and K. I. Bolotin, Nano Lett. **13**, 3626 (2013).
[50] J. Xia *et al.*, Nat. Phys. **41567**, 1005 (2020).
[51] C.-H. Chang, X. Fan, S.-H. Lin, and J.-L. Kuo, Phys. Rev. B **88**, 195420 (2013).
[52] L. Dong, A. M. Dongare, R. R. Namburu, T. P. O'Regan, and M. Dubey, Appl. Phys. Lett. **104**, 053107 (2014).
[53] W. Zhao, R. M. Ribeiro, M. Toh, A. Carvalho, C. Kloc, A. H. Castro Neto, and G. Eda, Nano Lett. **13**, 5627 (2013).